\documentclass{article}

\usepackage{arxiv}
\usepackage[utf8]{inputenc} 
\usepackage[T1]{fontenc}    
\usepackage{hyperref}       
\usepackage{url}            
\usepackage{booktabs}       
\usepackage{amsfonts}       
\usepackage{nicefrac}       
\usepackage{microtype}      
\usepackage{lipsum}
\usepackage{graphicx}
\usepackage{amsmath}
\usepackage{float}
\usepackage[compact]{titlesec}
\titlespacing{\section}{10pt}{*0}{*0}
\graphicspath{ {.} }

\title{Putting a price on tenure}

\author{
  Thiago Marzagão \\
}

\begin{document}
\maketitle

\begin{abstract}
Government employees in Brazil are granted tenure after three years on the job. Firing a tenured government employee is all but impossible, so tenure is a big employee benefit. But exactly how big is it? In other words: how much money is tenure worth to a government employee in Brazil? No one has ever attempted to answer that question. I do that in this paper. I use a modified version of the Sharpe ratio to estimate what the risk-adjusted salaries of government workers should be. The difference between actual salary and risk-adjusted salary gives us an estimate of how much tenure is worth to each employee. I find that in the 2005-2019 period the monthly value of tenure was R\$ 3980 to the median federal government employee, R\$ 1971 to the median state government employee, and R\$ 500 to the median municipal government employee.
\\
\\
Funcionários públicos brasileiros tornam-se estáveis três anos após a posse no cargo. Demitir um funcionário público estável é praticamente impossível, de modo que a estabilidade é um grande bônus para o funcionário. Mas exatamente quão grande? Em outras palavras: quanto dinheiro vale a estabilidade para um funcionário público no Brasil? Até hoje não se tentou responder essa questão. Faço isso neste paper. Usei uma versão modificada do índice de Sharpe para estimar qual deveria ser o salário risco-ajustado dos funcionários públicos. A diferença entra o salário real e o salário risco-ajustado nos dá uma estimativa do quanto vale a estabilidade para cada empregado.  Encontrei que o valor mensal da estabilidade no período 2005-2019 foi de R\$ 3980 para o funcionário público federal mediano, R\$ 1971 para o estadual e R\$ 500 para o municipal.
\end{abstract}


\vspace{12pt}
\section*{Introduction}

How much is tenure worth?\footnote{Institutional affiliation of the author: Data Intelligence Unit / Office of the Comptroller-General / Government of Brazil. Setor de Autarquias Sul (SAS), Quadra 1, Bloco A, Edifício Darcy Ribeiro, Brasília-DF  CEP: 70070-905, Brazil. Contact and ORCID information: thiago.marzagao@cgu.gov.br (https://orcid.org/0000-0003-0395-3985). All views expressed here are the authors' and do not represent views of any institutions he is affiliated with.}\footnote{Disclosure statement: the author is a tenured employee of the Brazilian federal government.}\footnote{I thank Leonardo Monasterio for circulating an earlier draft. I thank Arthur Kernkraut, Daniel Ortega, Paulo Antonacci, Raphael Cunha, Carlos Ramos, Simone Cuiabano, Daniel Duque, Thársis Souza, Silvia Wada, and Francisco da Costa for helpful comments. All errors are mine.}

Government employees in Brazil are granted tenure after three years on the job (Federal Constitution of Brazil, art. 41). Firing a tenured government employee is all but impossible, so tenure is a big employee benefit. But exactly how big is it? In other words: if you are a government employee in Brazil, how much is tenure worth to you? No one has attempted to answer that question before.

In this paper I put a price on tenure. I calculate how much tenure is worth, in R\$, to each employee of a sample of 188847 thousand government employees. I then analyze how the monetary value of tenure varies across government levels (federal, state, municipal).

Importantly, this paper is \emph{not} about the wage gap between public and private sector. It is well known that government employees are overpaid. For instance, in Brazil federal government employees make 96\% more than private sector employees of the same gender, race, age, working time, and schooling.\cite{worldbank} And that is before taking into account selection biases; the workers who self-select into government would make \emph{less} money in the private sector than their demographically similar counterparts do.\cite{araujo} The public-sector wage premium is of crucial importance, as it can affect occupational choices and hurt productivity.\cite{10.1093/jeea/jvaa038} That is not the topic of this paper though. I am not trying to estimate the public-sector wage premium. In this paper I merely try to come up with a monetary value for tenure.

\vspace{12pt}
\section*{Basic idea}
We generally expect returns to be higher for riskier assets (like stocks) and lower for safer assets (like bonds). By "risk" economists usually mean the standard deviation of the asset's price over a certain period.\footnote{That measure captures "ordinary" risk, but not "black swan" events like pandemics and wars.\cite{taleb}} By "return" economists mean the change in the asset's price over a certain period.\footnote{Risk and return are each a function of the period we choose. For instance, as \cite{siegel} notes, "as the holding period increases to between fifteen and twenty years, the riskiness of stocks falls bellow that of fixed-income assets" (p. 245).} The relationship between risk and return can thus be expressed in a number of ways, the most popular being the Sharpe ratio\cite{sharpe}:

\begin{center}
    $S_{a} = \dfrac{R_{a} - R_{f}}{\sigma_{a}} $
\end{center}

where $S_{a}$ is the Sharpe ratio of asset $a$, $R_{a}$ is the return rate of asset $a$, $R_{f}$ is the return rate of a risk-free asset (often set to zero), and $\sigma_{a}$ is the standard deviation of asset $a$'s price.

The Sharpe ratio allows us to compare different assets. Imagine two assets, $a$ and $b$, with standard deviations 0.1 and 0.2 respectively, and the same return rate, 0.3. If we set the risk-free return rate ($R_{f}$) to zero then $S_{a}$ is 3 and $S_{b}$ is 2. This means that asset $a$ gives us more return per unit of risk than asset $b$.

We can extend that logic to salaries. Salaries, like stock prices, can go up and down. The risk of your salary going down is much lower in the public sector than in the private sector. A tenured employee rarely experiences a salary reduction unless he or she gets fired, which is all but impossible (in the 2003-2019 period fewer than 0.5\% of the employees in the federal civil service were fired)\footnote{https://www1.folha.uol.com.br/mercado/2020/01/nenhum-dos-7766-servidores-expulsos-desde-2003-saiu-por-mau-desempenho.shtml}. Since public-sector salaries have less risk, they should (all else equal) be lower.

That is the basic idea of this paper. I treat salaries more or less like financial assets, with a return rate and a risk rate, and then I use (a modified version of) the Sharpe ratio to calculate what public-sector wages should be if they were risk-adjusted. The difference between actual public-sector wages and risk-adjusted public-sector wages is a measure of how much tenure is worth. The next two sections explain in detail how these estimates were produced.

\vspace{12pt}
\section*{Data}

I got all the data from RAIS (\emph{Relação Anual de Informações Sociais}), a database maintained by the federal government. RAIS contains employment data for workers who are formally employed in Brazil, be it in the private or public sector. For every employee RAIS contains monthly wages, employer ID, and a myriad of other data. I used RAIS data from 2005 (the first year for which full employer data are available) to 2019 (the latest year for which RAIS data are available).

Importantly, RAIS does not have any data on informal workers, self-employed workers, business owners, and some company officers \footnote{Some company officers are contracted as service providers with a fixed mandate. They are not considered employees and therefore are not required to be on RAIS}. That means I am underestimating the riskiness of private sector work, as the incomes of informal workers and self-employed workers certainly fluctuate more than the incomes of formally employed workers. That cannot be helped; even if I had access to every worker's tax returns I would still not have income data on informal workers, and I would only have yearly income data on self-employed workers (as opposed to the monthly RAIS data).

Due to limitations in computing power I could not use the entire RAIS. Instead I randomly sampled, for each of the following categories, 100 thousand people: private sector employees; tenure-track federal government employees; tenure-track state government employees; and tenure-track municipal government employees. That yielded a total of 400 thousand people. By "tenure-track" I mean both tenured government employees and "pre-tenure" government employees, i.e., those employees who are still in the three-year probationary period. Practically all pre-tenure government employees eventually become tenured, so for simplicity I will consider them as tenured employees and refer to them as such.

I sampled those 400 thousand people from RAIS 2005, the first year for which RAIS has sufficient employer information (prior to 2005 the "employer nature" field - which tells us whether the employer is a government organization, a private company, a non-profit, etc - followed a different logic). I then checked, for each year in the 2005-2019 period (2019 is the last year for which RAIS data are available), whether any of the sampled employees switched categories in that year - i.e., whether any private sector workers became government workers, and whether any government workers became private sector workers or switched to a different government level (federal to state, municipal to federal, etc). I discarded all people who had switched categories at any point. That left a total of 273094 people, distributed as follows:\footnote{In some cases people disappear from RAIS for a time. For example, an employee who was fired in 2007, became self-employed, and was hired again in 2010 will not be on RAIS in the 2008-2009 interval. I could drop such cases and keep only people who were on RAIS 2005-2019 uninterruptedly. But that would drop precisely the people who are most at risk of losing their jobs. So instead I keep all 274690 people and I concatenate all the monthly data points I have for each of them (which results in time series of different lengths for different people; for instance, someone who was on RAIS for the whole 2005-2019 interval will have a time series of 180 data points, but someone who was not on RAIS in 2008 and 2009 will have a time series of only 156 data points).}

\begin{table}[h]
 \caption{sample distribution}
  \centering
  \begin{tabular}{rc}
    \toprule
    private sector & 84247 \\
    federal government & 69428 \\
    state government & 61634 \\
    municipal government & 57785 \\
    \hline
    total & 273094 \\
    \bottomrule
  \end{tabular}
  \label{tab:table1}
\end{table}

\vspace{12pt}
\section*{Method}

The first step was to calculate a modified version of the Sharpe ratio - called the Sortino ratio - for each of the 84247 private sector employees in the sample, for the 2005-2019 interval.

The (unmodified) Sharpe ratio treats ups and downs equally. If a stock price or salary goes up this increases the price's standard deviation, which decreases the Sharpe ratio. That is often not what we want; whether we are talking about stock prices or wages, the risk we are interested in is the downside risk, not the upside risk. The Sortino ratio\cite{sortino} solves this problem by replacing the standard deviation with the downside deviation (also called the lower semi-deviation), which only captures negative fluctuations. Here this means the following:

\begin{center}
    $S_{i}' = \dfrac{R_{i} - R_{f}}{DD_{i}} $
\end{center}

where $S_{i}'$ is the Sortino ratio of employee $i$'s wages; $R_{i}$ is the return rate of employee $i$'s wages, which I define as $((k + \sum_{i=0}^{n} w_{i})/k)-1$, where $k$ is $100 (\sum_{i=0}^{2} w_{i}) / 3)$, a proxy for how much $i$ invested in her human capital,\footnote{$100 (\sum_{i=0}^{2} w_{i}) / 3)$ is largely arbitrary, I just needed a number big enough to avoid Sortino ratios of thousands or millions.} and $\sum_{i=0}^{n} w_{i}$ is the sum of all wages\footnote{All wages were adjusted to December 2019 values according to the IPCA inflation index.} $i$ received between Jan/2005 and Dec/2019;\footnote{That does not mean that every employee received a wage every month between Jan/2005 and Dec/2019. As mentioned before, the length of the time series is different for different employees, as people can become unemployed, self-employed, etc.} $R_{f}$ is the return rate of a risk-free asset, which here I set to zero; and $DD_{i}$ is the downside deviation of employee $i$'s monthly wages between Jan/2005 and Dec/2019, based on the changes from one month to the next.\footnote{The downside deviation is calculated by summing up all the negative fluctuations, squaring each of them, summing up the squares, dividing that sum by the total number of observations (both positive and negative fluctuations), and taking the square root of that fraction.}

The second step was to calculate the mean Sortino ratio of every income bracket of the 84247 private sector employees. I could have used deciles to create those brackets. But the top decile has a range that is too wide. In 2005, for instance, the top decile comprehends salaries between R\$ 3 thousand and R\$ 154 thousand (in December 2019 R\$).

So instead of deciles I used Jenks natural breaks, which is a clustering algorithm for 1-dimensional data (the Jenks algorithm minimizes each cluster's mean deviation from the cluster mean and maximizes each cluster's deviation from the means of the other clusters).\cite{jenks}

The result was the following income brackets and corresponding mean Sortino ratios (based on 2005 salaries in December 2019 R\$):

\begin{table}[h]
 \caption{mean Sortino ratios, per income bracket}
  \centering
  \begin{tabular}{lc}
    \toprule
    income bracket & mean Sortino ratio \\
    \midrule
    (R\$ 2, R\$ 1079( & 23.03 \\
    (R\$ 1079, R\$ 2185( & 13.37 \\
    (R\$ 2185, R\$ 4101( & 13.00 \\
    (R\$ 4101, R\$ 7115( & 13.50 \\
    (R\$ 7115, R\$ 11593( & 13.41 \\
    (R\$ 11593, R\$ 18087( & 13.21 \\
    (R\$ 18087, R\$ 28369( & 11.62 \\
    (R\$ 28369, R\$ 46357( & 10.54 \\
    (R\$ 46357, R\$ 72539( & 7.34 \\
    (R\$ 72539, R\$ 154022( & 5.27 \\
    \bottomrule
  \end{tabular}
  \label{tab:table2}
\end{table}

The third step was to compute the Sortino ratio for each of the 188847 government employees in the sample, for the 2005-2019 interval.

The fourth step was to find, for each government employee, the mean Sortino ratio of the corresponding private sector income bracket.\footnote{The idea is to "match" - in a very loose sense of the word - each government employee to a comparable set of private-sector employees. I do not need an exact match - in terms of schooling, work experience, etc -, as I am not trying to estimate the public-sector wage premium.} I then used that information to find out, for every government employee, what her return rate should have been in order to equalize her Sortino ratio with the average Sortino ratio of the corresponding income bracket in the private sector.\footnote{There are government salaries higher than R\$ 154022. They have no equivalent private sector bracket (private sector salaries only go up to R\$ 154022 in 2005). In these cases we used the average Sortino rate of the top income bracket.}

A concrete example may help. Imagine a government employee - let us call her Jane - who received a total of R\$ 300 thousand in wages between Jan/2005 and Dec/2019, and whose first three monthly wages were R\$ 4000, R\$ 5000, and R\$ 6000. That gives us a $k$ (our proxy for human capital) of $100((4000 + 5000 + 6000)/3)=500000$ and a return rate ($R_{i}$) of $((500000 + 600000)/500000) - 1 = 0.6$, i.e., 60\%.  Assume that the downside deviation of Jane's wages in that same period was 0.025. That results in a Sortino ratio of 24 ($0.6/0.025=24$).

The private sector income bracket that contains R\$ 15000 is the one that goes from R\$ 11593 to R\$ 18087 (see Table 2 above). That income bracket has a mean Sortino ratio of 13.21. Keeping Jane's downside deviation constant, what should her return rate have been to produce a Sortino ratio of 13.21? The answer is $13.21(0.025) = 0.33$. That means Jane should have received a total of R\$ 99075 in wages between Jan/2005 and Dec/2019 - and not the R\$ 300 thousand she actually received; R\$ 99075 is the amount that would have equalized her Sortino ratio with the Sortino ratio of the corresponding private-sector income bracket.

The difference between those two values, $300000-99075=200925$, is \emph{the total value of tenure to Jane between Jan/2005 and Dec/2019.} In other words, tenure was worth a total of R\$ 200925 to Jane between Jan/2005 and Dec/2019. If we want a monthly value we simply divide that total by the number of months Jane worked in that period. Suppose she worked every month. In that case tenure was worth about R\$ 1116.25 to Jane every month ($200925/180=1116.25$).

\vspace{12pt}
\section*{Salaries as both dividends and prices}

The method described above takes some liberties with the Sortino ratio. It treats the wages themselves as the returns, but it treats the monthly variation from one month to the next as the basis for the downside deviation. In other words, the numerator treats wages as dividends but the denominator treats wages as stock prices.

That inconsistency is unavoidable. If the numerator treated wages as stock prices then the return would be based solely on the difference between the first and last wages in the time series. That would discard all the information contained in the wages in-between - i.e., almost all of the data. Conversely, if the denominator treated wages as a return on $k$ then there would be no negative variation, as salaries cannot be negative, whereas in the real world it does matter that a person sees her income diminish from one month to the next and we want to capture that somehow.

But that inconsistency is not important here, as I am not interested in the Sortino ratios themselves. I am merely using the Sortino ratios as a mechanism to compare the risk-adjusted returns of private- and public-sector workers. Jane's Sortino ratio fo 24 is irrelevant in itself - that figure could be 2 or 200. What matters is Jane's Sortino ratio in comparison to the Sortino ratio of her corresponding private-sector income bracket - 13.21. In other words, what matters is that I am computing the Sortino ratios the same way for both public- and private-sector workers.

\vspace{12pt}
\section*{Results}

The monthly value of tenure is R\$ 3980 to the median federal worker, R\$ 1971 to the median state worker, and R\$ 500 to the median municipal worker. The histograms below show the distribution of tenure values for each government level (in each case I removed the bottom 2.5\% and the top 2.5\% values, to avoid distorting the visualization).

\begin{figure}[htp]
  \centering
  \includegraphics[scale=0.75]{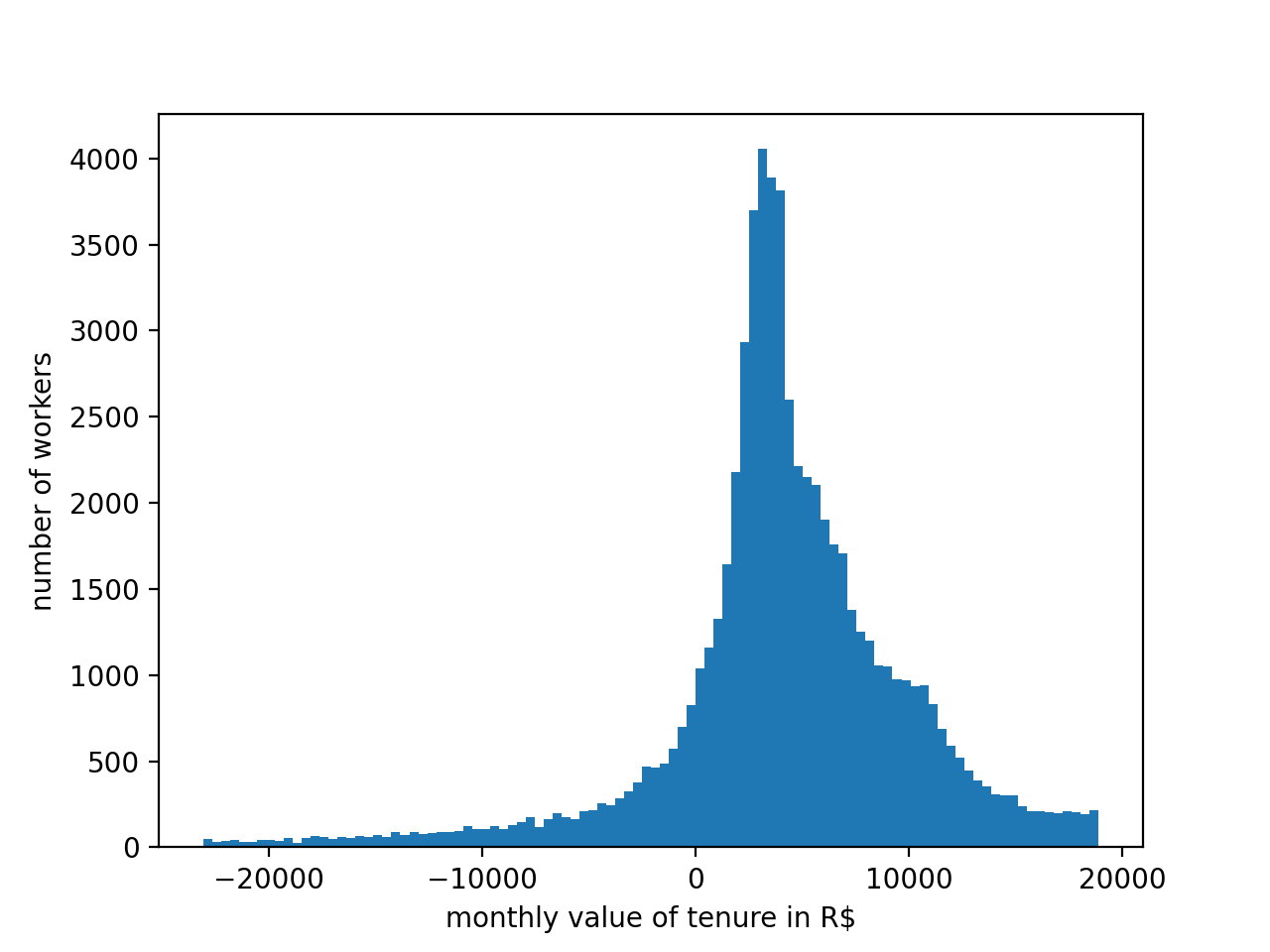}
  \caption{monthly value of tenure to the median federal worker}
  \label{fig:fed}
\end{figure}

\pagebreak

\begin{figure}[htp]
  \centering
  \includegraphics[scale=0.75]{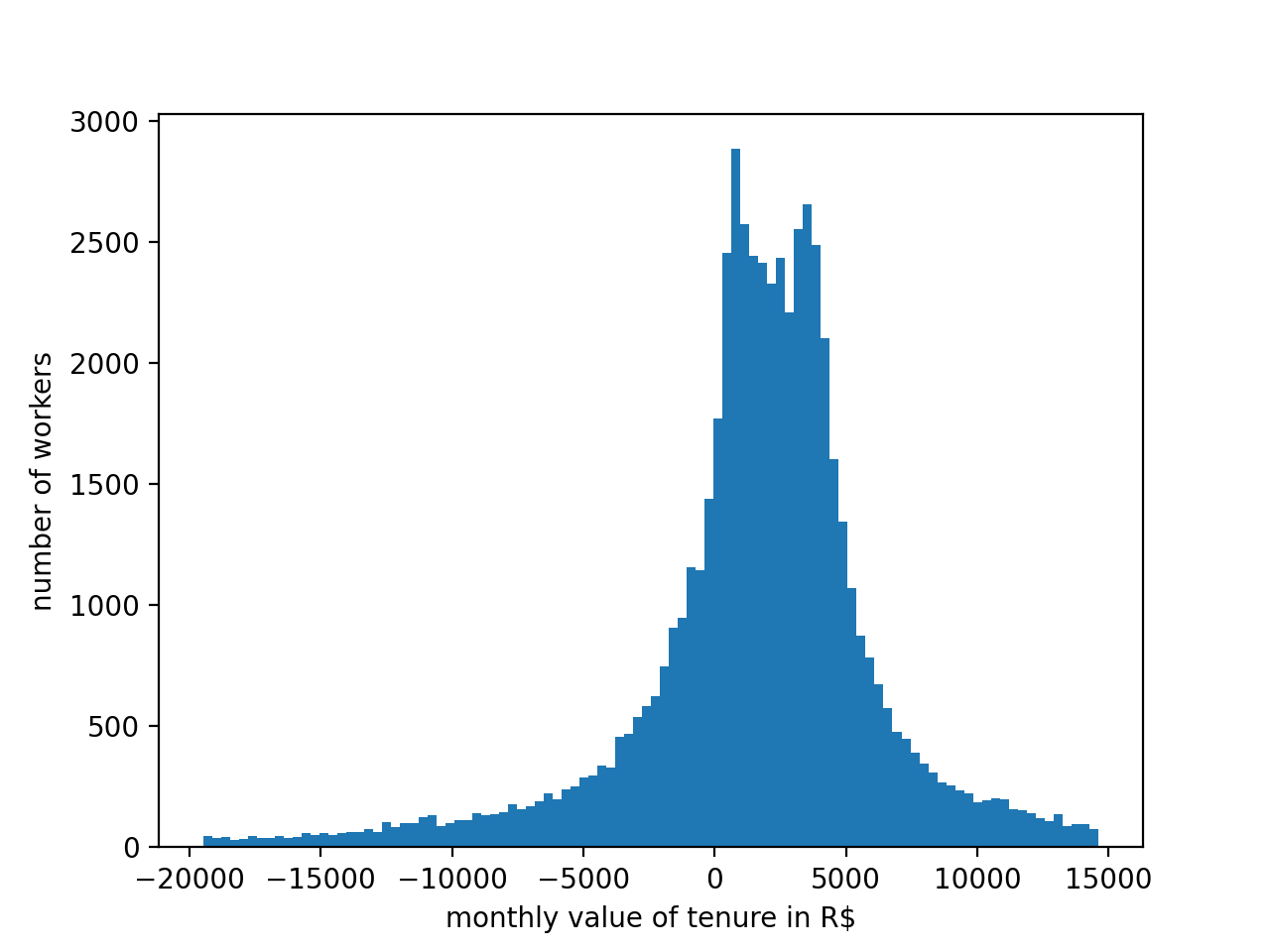}
  \caption{monthly value of tenure to the median state worker}
  \label{fig:est}
\end{figure}

\begin{figure}[!htp]
  \centering
  \includegraphics[scale=0.75]{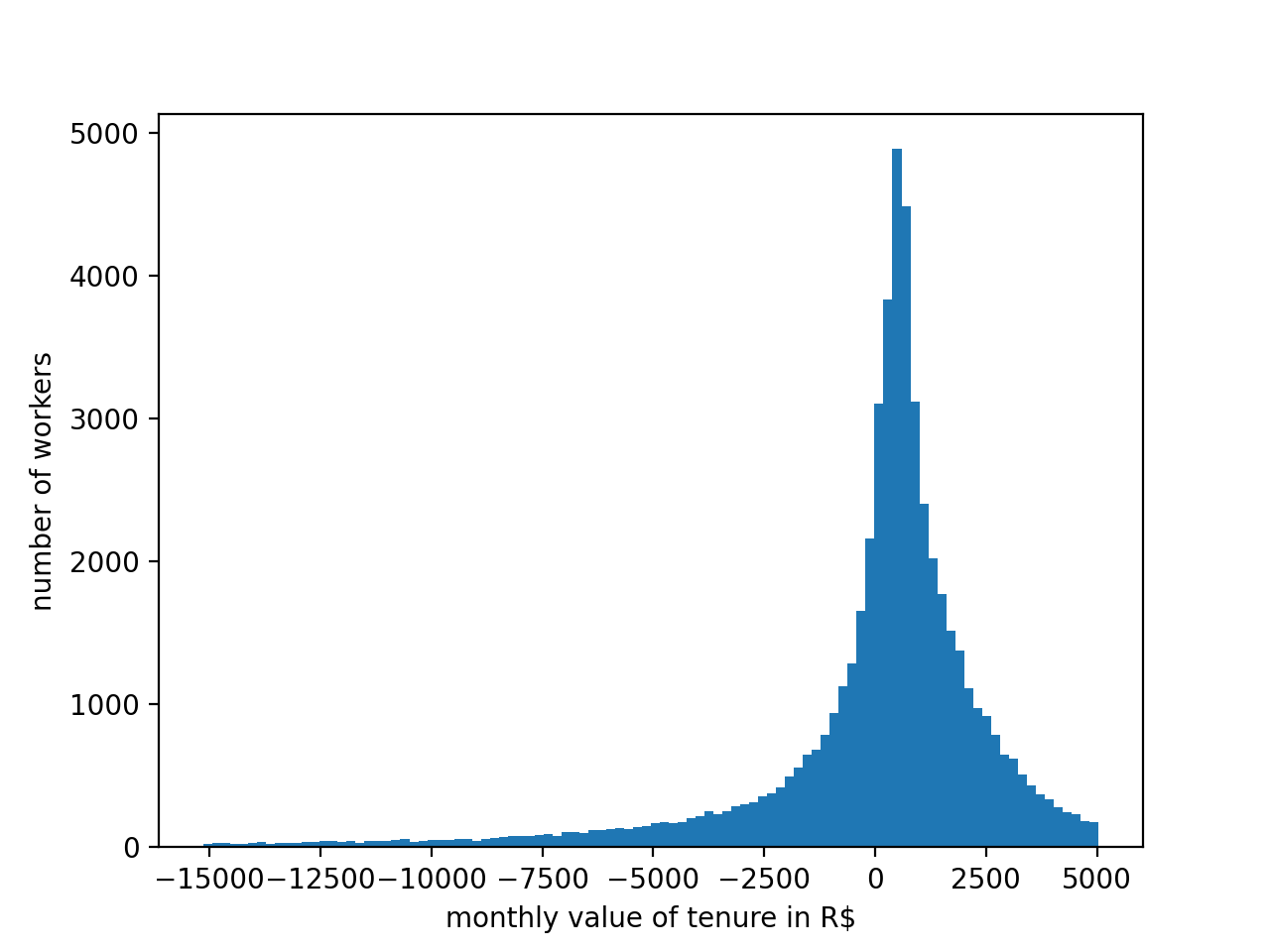}
  \caption{monthly value of tenure to the median municipal worker}
  \label{fig:mun}
\end{figure}

\pagebreak

Interestingly, as the histograms show, to some government workers tenure has a negative value. From a risk-reward perspective those workers might be better off in the private sector: they might face a higher chance of being fired, but the extra risk would be more than compensated by the higher salary returns. Such cases are not the norm though: they represent 15\% of the federal workers, 26\% of the state workers, and 33\% of the municipal workers.

To the vast majority of government workers, tenure is worth a considerable, positive sum. For the median federal worker tenure is worth a sum equivalent to 45\% of her salary. For the median state and municipal workers that figure is 43\% and 24\% respectively. That means the public debate about civil service compensation is missing an important component, as current estimates of the public-sector wage premium (e.g. \cite{worldbank}, \cite{araujo}) are based on salaries alone.

\vspace{12pt}
\section*{Conclusion}

This paper brings the notion of risk-adjusted returns to the research on public-sector compensation. I hope that this contribution will help inform the public debate and policy-making on the subject.

The approach taken here could be refined and extended in a number of ways. Here I am only interested in the value of tenure, not in the public-sector wage premium. A natural next step would be to do both estimations simultaneously. That would require surveying private- and public-sector employees for a number of years. RAIS does not have enough data on education, work experience, etc, and the main alternative - IBGE's PNAD (\emph{Pesquisa Nacional por Amostra de Domicílios}) - has that data but it does not track the same people over time. That data collection effort might be worth it, especially if it allows us to incorporate categories that are not on RAIS (informal workers, self-employed workers, etc).

Another direction would be to estimate the price of tenure in completely different ways and then compare the estimates with the ones found here. For instance, we could ask a sample of government employees how much additional \$ they would require in order to give up tenure. That answer, biased as it might be (employees might inflate their true preferences when answering the survey) could let us estimate the gap between actual risk and perceived risk.

\bibliographystyle{unsrt}  

\bibliography{references}

\end{document}